\documentclass[letterpaper,english,aps,prl,letterpaper,twocolumn,showpacs,floatfix,groupedaddress,superscriptaddress]{revtex4}
\usepackage{mathptmx}
\usepackage[T1]{fontenc}
\usepackage[latin9]{inputenc}
\usepackage{amsmath}
\usepackage{graphicx}
\usepackage{amssymb}

\makeatletter

\providecommand{\tabularnewline}{\\}

\usepackage{babel}
\makeatother

\begin{document}

\title{First-Principles Semiclassical Initial Value Representation Molecular
Dynamics \footnote{On the anniversary of the 100th year of foundation of the "Società Chimica Italiana"}}

\author{Michele Ceotto}

\address{Dipartimento di Chimica Fisica ed Elettrochimica, Università di Milano,
via Golgi 19, 20133 Milano, Italy}

\email{michele.ceotto@unimi.it, aspuru@chemistry.harvard.edu}

\author{Sule Atahan}

\address{Department of Chemistry and Chemical Biology, Harvard University,
12 Oxford Street, 02138, Cambridge, MA}

\author{Sangwoo Shim}

\address{Department of Chemistry and Chemical Biology, Harvard University,
12 Oxford Street, 02138, Cambridge, MA}

\author{Gian\ Franco\ Tantardini}

\address{Dipartimento di Chimica Fisica ed Elettrochimica, Università di Milano,
via Golgi 19, 20133 Milano, Italy}

\affiliation{Istituto CNR di Scienze e Tecnologie Molecolari, via Golgi 19, 20133
Milano, Italy}

\author{Al\'an Aspuru-Guzik}

\address{Department of Chemistry and Chemical Biology, Harvard University,
12 Oxford Street, 02138, Cambridge, MA}

\email{aspuru@chemistry.harvard.edu}

\begin{abstract}
In this work, we explore the use of the semiclassical initial value
representation (SC-IVR) method with first-principles electronic structure
approaches to carry out classical molecular dynamics. The proposed
approach can extract the vibrational power spectrum of carbon dioxide
from a single trajectory providing numerical results that agree with
experiment and quantum calculations. The computational demands of
the method are comparable to those of classical single-trajectory
calculations, while describing uniquely quantum features such as the
zero-point energy and Fermi resonances. The method can also be used
to identify symmetry properties of given vibrational peaks and investigate
vibrational couplings by selected classical trajectories. The accuracy
of the method degrades for the reproduction of anharmonic shifts for
high-energy vibrational levels. 
\end{abstract}


\maketitle

\section{Introduction}

Algorithms for the simulation of molecular dynamics belong to the
fundamental toolset of modern theoretical chemical physics. Classical
simulation methods are able to study systems with up to millions of
particles but are unable to describe quantum effects such as tunelling
and delocalization. Exact quantum mechanical methods are restricted
to a few quantum particles, especially when pre-computed analytical
potential energy surfaces (PES) are employed. 

First-principles molecular dynamics (FPMD) algorithms have been introduced
as an alternative to the pre-calculation of the PES. FPMD avoids any
source of error originated from the fitting of the PES. This is particularly
true for many degrees of freedom, where the fitting procedure might
not represent the many-dimensional surface accurately. In this family
of methods, the potential and its derivatives are calculated \emph{on-the-fly}
as the dynamical simulation progresses and are directly obtained from
electronic structure calculations. In the Born-Oppenheimer molecular
dynamics (BOMD) approach, the electronic structure calculations for
a given simulation step are converged based on previous step information.
This approach can lead to systematic energy drifts and several methods
have been proposed to avoid this effect \citep{Herbert_acceleratedBOMD}.
Alternatively, extended Lagrangian molecular dynamics approaches (ELMD)
\citep{CPMD,Schlegel_Voth_gaussian,Herbert_curvysteps,Tuckerman_grid}
involve the propagation of nuclear and electronic degrees of freedom
simultaneously. The electronic degrees of freedom are assigned to
classical variables that are propagated using classical equations
of motion and these can be expanded in terms of plane waves \citep{CPMD},
Gaussian functions \citep{Herbert_curvysteps} or real-space grids
\citep{Tuckerman_grid}. Usually ELMD propagation is computationally
more efficient, however questions have raised on whether the resultant
energy surface remains close to the actual Born-Oppenheimer one and
about disturbing dependencies on the fictitious electronic masses
\citep{Herbert_curvysteps,Tangey}.

While the evaluation of the potential \emph{on-the-fly} can be easily
integrated with classical simulations, the delocalized nature of quantum
mechanical propagation has led to the development of many alternative
approaches for the simulation of quantum dynamics. For example, the
path-integral centroid molecular dynamics approach \citep{Voth_AICMD_Pavese}
includes quantum nuclear effects employing an extended Lagrangian.
Alternatively, in the variational multi-configuration Gaussian wavepacket
method (vMCG) \citep{Burghardt_worthreview} the quantum wavepackets
are represented by fixed-width Gaussian functions for which the potential
is approximated to be locally harmonic. Other approaches introduce
a mean field approximation and then update the dynamics in a time-dependent
self-consistent fashion \citep{Iyengar,Jungwirth}.

Semiclassical molecular dynamics methods \citep{Miller_avd_74,Miller_JPC_featurearticle,Heller_frozengaussian,Heller_review,Coker,Batista,Grossmann,Manolopoulos,Miller_vari,Pollak}
are based on classical trajectories and therefore are amenable for
carrying out \emph{on-the-fly} calculation of the potential. The benefits
of calculating the potential only when needed have been suggested
by Heller and co-workers \citep{Heller_review,HellerVanVoorhis}.
In between formally exact quantum methods and classical dynamics,
semi-classical methods include quantum effects approximately. Two
representative semi-classical approaches are the coupled coherent
states (CCS) technique \citep{Shalashilin_review} and the ab initio
multiple spawing method (AIMS) algorithm \citep{Martinez_reviewAIMS}.
In the CCS approach, several grids of coherent states are classically
propagated and their trajectories can be derived from first principle
dynamics. In AIMS, the nuclear wavefunction are spawned onto a multiple
potential surface basis set. This set is made of adaptive time-dependent
fixed-width Gaussian functions, which are generated by classical Newtonian
dynamics.

\section{First-Principles SC-IVR}

In this work, we show how the semiclassical initial value representation
(SC-IVR) \citep{Miller_JPC_featurearticle} method can be coupled
tightly and naturally, without any mayor change in formulation, with
first principles electronic structure approaches to carry out classical
molecular dynamics. We show how the method is able to reproduce approximately
quantum effects such as the vibrational power spectra using a single,
short classical trajectory using computational resources comparable
to those employed in first-principles molecular dynamics calculations.
Calculations employing multiple trajectories can in principle be more
accurate (and more computational intense as well), but here we focus
on analyzing the predictive power of single trajectory runs. Finally,
we describe how different approaches can be used in conjunction with
this method for studying the symmetry of the vibrational states either
by arranging the initial conditions of the classical trajectory or
by employing the symmetry of the coherent state basis. 

In the SC-IVR method, the propagator in $F$ dimension is approximated
by the phase space integral,

\begin{eqnarray}
e^{-i\hat{H}t/\hbar}= & \frac{1}{\left(2\pi\hbar\right)^{F}}\int d\mathbf{p}\left(0\right)\int d\mathbf{q}\left(0\right)\: C_{t}\left(\mathbf{p}\left(0\right),\mathbf{q}\left(0\right)\right)\nonumber \\
 & e^{iS_{t}\left(\mathbf{p}\left(0\right),\mathbf{q}\left(0\right)\right)/\hbar}\left|\mathbf{p}\left(t\right),\mathbf{q}\left(t\right)\left\rangle \right\langle \mathbf{p}\left(0\right),\mathbf{q}\left(0\right)\right|\label{eq:SCTime_evolution}\end{eqnarray}

where $\left(\mathbf{p}\left(t\right),\mathbf{q}\left(t\right)\right)$
are the set of classically-evolved phase space coordinates, $S_{t}$
is the classical action and $C_{t}$ is a pre-exponential factor.
In the Heller-Herman-Kluk-Kay \citep{Heller_frozengaussian,HermanLukcoherstates}
version of the SC-IVR, the prefactor involves mixed phase space derivatives
\begin{eqnarray}
C_{t}\left(\mathbf{p}\left(0\right),\mathbf{q}\left(0\right)\right) & =\label{eq:prefactor}\\
\sqrt{\frac{1}{2}\left|\frac{\partial\mathbf{q}\left(t\right)}{\partial\mathbf{q}\left(0\right)}+\frac{\partial\mathbf{p}\left(t\right)}{\partial\mathbf{p}\left(0\right)}-i\hbar\gamma\frac{\partial\mathbf{q}\left(t\right)}{\partial\mathbf{p}\left(0\right)}+\frac{i}{\gamma\hbar}\frac{\partial\mathbf{p}\left(t\right)}{\partial\mathbf{q}\left(0\right)}\right|}\nonumber \end{eqnarray}
as well as a set of reference states $\left\langle \mathbf{q}\left|\right.\mathbf{p}\left(t\right),\mathbf{q}\left(t\right)\right\rangle =\prod_{i}\left(\mathbf{\gamma}_{i}/\pi\right)^{F/4}\mbox{exp}\left[-\mathbf{\gamma_{i}}\cdot\left(q_{i}-q_{i}\left(t\right)\right)/2+ip_{i}\left(t\right)\cdot\left(q_{i}-q_{i}\left(t\right)\right)/\hbar\right]$
of fixed width $\gamma_{i}$. For bound systems, the widths are usually
chosen to match the widths of the harmonic oscillator approximation
to the wave function at the global minimum and no significant dependency
has been found under width variation \citep{Miller_vari}. By introducing
a $2F\times2F$ symplectic (monodromy) matrix $\mathbf{M}\left(t\right)\equiv\partial\left(\left(\mathbf{p}_{t},\mathbf{q}_{t}\right)/\partial\left(\mathbf{p}_{0},\mathbf{q}_{0}\right)\right)$,
one can calculate the pre-factor of Eq. (\ref{eq:prefactor}) from
blocks of $F\times F$ size and monitor the accuracy of the classical
approximate propagation by the deviation of its determinant from unity.
Wang \emph{et al.} suggested calculating the determinant of the positive-definite
matrix $\mathbf{M}^{T}\mathbf{M}$ instead \citep{Wang_Dmatrix} and
we monitored the same quantity for this work. The spectral density
is obtained as a Fourier transform of the surviving probability\cite{Heller_frozengaussian}.
The SC-IVR expression of the probability of survival for a phase-space
reference state $\left|\chi\right\rangle =\left|p_{N},q_{N}\right\rangle $
is

\begin{widetext}

\begin{equation}
\left\langle \chi\left|e^{-i\hat{H}t/\hbar}\right|\chi\right\rangle =\frac{1}{\left(2\pi\hbar\right)^{F}}\int d\mathbf{p}\left(0\right)\int d\mathbf{q}\left(0\right)\: C_{t}\left(\mathbf{p}\left(0\right),\mathbf{q}\left(0\right)\right)e^{iS_{t}\left(\mathbf{p}\left(0\right),\mathbf{q}\left(0\right)\right)/\hbar}\left\langle \chi\left|\right.\mathbf{p}\left(t\right),\mathbf{q}\left(t\right)\right\rangle \left\langle \mathbf{p}\left(0\right),\mathbf{q}\left(0\right)\left|\right.\chi\right\rangle .\label{eq:SC_prop_matrix_elemts}\end{equation}

\end{widetext}

The phase-space integral of Eq. (\ref{eq:SC_prop_matrix_elemts})
is usually computed using Monte Carlo methods. If the simulation time
is long enough, the phase space average can be well approximated
by a time average integral. This idea has been suggested and implemented
by Kaledin and Miller \citep{Alex_Mik} to obtain the TA (Time Averaging
\citep{TA-Kay}) SC-IVR approximation for the spectral density,

\begin{widetext}

\begin{eqnarray}
I\left(E\right) & = & \frac{1}{\left(2\pi\hbar\right)^{F}}\int d\mathbf{p}\left(0\right)\int d\mathbf{q}\left(0\right)\frac{\mbox{Re}}{\pi\hbar T}\int_{0}^{T}dt_{1}\int_{t_{1}}^{T}dt_{2}\: C_{t_{2}}\left(\mathbf{p}\left(t_{1}\right),\mathbf{q}\left(t_{1}\right)\right)\nonumber \\
 & \times & \left\langle \chi\left|\right.\mathbf{p}\left(t_{2}\right),\mathbf{q}\left(t_{2}\right)\right\rangle e^{i\left(S_{t_{2}}\left(\mathbf{p}\left(0\right),\mathbf{q}\left(0\right)\right)+Et_{2}\right)/\hbar}\left[\left\langle \chi\left|\right.\mathbf{p}\left(t_{1}\right),\mathbf{q}\left(t_{1}\right)\right\rangle e^{i\left(S_{t_{1}}\left(\mathbf{p}\left(0\right),\mathbf{q}\left(0\right)\right)+Et_{1}\right)/\hbar}\right]^{*}\label{eq:TA_spectrdens}\end{eqnarray}

\end{widetext}where $\left(\mathbf{p}\left(t_{1}\right),\mathbf{q}\left(t_{1}\right)\right)$
and $\left(\mathbf{p}\left(t_{2}\right),\mathbf{q}\left(t_{2}\right)\right)$
are variables that evolve from the same initial conditions but to
different times, and $T$ is the total simulation time. The advantage
of this approach is that the additional time integral can in principle
replace the need for phase-space averaging in the large-time limit
of a single trajectory. Calculations of the vibrational spectra of
systems such as the water molecule have proved to be very accurate
using the TA-SC-IVR approach and its inexpensive single-trajectory
variant showed significant improvements over the simple harmonic approximation
for excited vibrational levels \citep{Alex_Mik}. In order to make
Eq. (\ref{eq:TA_spectrdens}) less computationally demanding, one
can employ the separable approximation \citep{Alex_Mik}, where the
pre-factor of Eq. (\ref{eq:TA_spectrdens}) is approximated as a phase,
$C_{t_{2}}\left(\mathbf{p}\left(t_{1}\right),\mathbf{q}\left(t_{1}\right)\right)=\mbox{Exp}\left[i\left(\phi\left(t_{2}\right)-\phi\left(t_{1}\right)\right)/\hbar\right],$
and $\phi\left(t\right)=\mbox{phase}\left[C_{t}\left(\mathbf{p}\left(0\right),\mathbf{q}\left(0\right)\right)\right]$.
Using this approximation, Eq. (\ref{eq:TA_spectrdens}) becomes\begin{eqnarray}
I\left(E\right) & = & \frac{1}{\left(2\pi\hbar\right)^{F}}\frac{1}{2\pi\hbar T}\int d\mathbf{p}\left(0\right)\int d\mathbf{q}\left(0\right)\nonumber \\
 & \times & \left|\int_{0}^{T}dt\left\langle \chi\left|\right.\mathbf{p}\left(t\right),\mathbf{q}\left(t\right)\right\rangle \right.\label{eq:sep_approx}\\
 & \times & \left.e^{i\left(S_{t}\left(\mathbf{p}\left(0\right),\mathbf{q}\left(0\right)\right)+Et+\phi_{t}\left(\mathbf{p}\left(0\right),\mathbf{q}\left(0\right)\right)/\hbar\right)}\right|^{2}\nonumber \end{eqnarray}
leading to a simplification of the double-time integration to a single
time integral. The resulting integral is positive definite, making
more amenable for Monte Carlo integration. Our numerical tests show
that the results of carrying out this approximation are essentially
identical to the double time integral approach when using a single
trajectory. In this paper results will be reported by use of this
last approximation, since it is computationally cheaper and numerically
more stable than Eq. (\ref{eq:TA_spectrdens}). 

\begin{figure}
\begin{centering}
\includegraphics[scale=0.45]{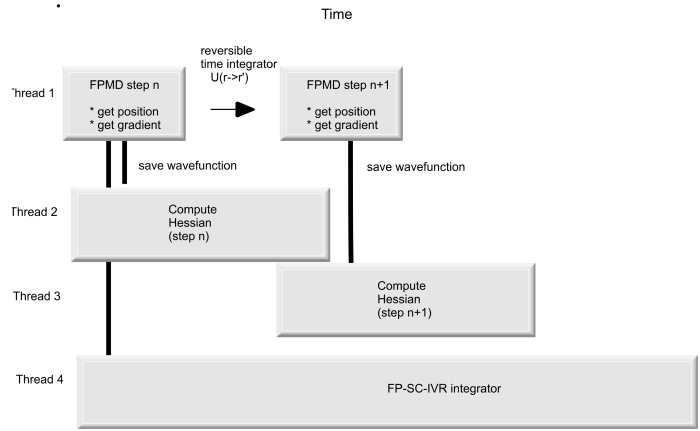}\caption{\label{fig:algorithm:}First-principles SC-IVR algorithm: At each
time step electronic wavefunction are saved to calculated nuclear
Hessian. Nuclear positions, gradients and Hessian are accumulated
for the spectral time-average integral.}

\par\end{centering}
\end{figure}

For this work, we compute the potential energy surface at each nuclear
configuration directly from the Kohn-Sham orbitals expanded on a non-orthogonal
Gaussian basis. Gradients and Hessians at each nuclear configuration
are obtained analytically from electronic orbitals. The evaluation
of the potential represents most of the computational effort of our
approach, which is roughly few hours of computer time using standard
desktop machines for a $1\:\mbox{cm}^{-1}$ spectrum resolution. The
nuclear equations of motion are \begin{equation}
M_{I}\ddot{\mathbf{R}}_{I}=-\nabla_{I}\begin{array}{c}
\textrm{min}\\
\mathbf{C}\end{array}E_{DFT}\left[\mathbf{C},\mathbf{R}_{I}\right]\label{eq:BOMD}\end{equation}
where $\mathbf{C}$ is the rectangular matrix of the lowest occupied
orbitals and the classical propagation is performed according to the
velocity-Verlet algorithm, as implemented in the Q-Chem package \citep{QChem}.
At each time step, the potential, nuclear gradient and Hessian are
used to calculate the action, pre-factor and coherent state overlaps
necessary for the TA-SC-IVR method (Eqs. \ref{eq:TA_spectrdens} and
\ref{eq:sep_approx}). A schematic representation of an implementation
of the algorithm for a multithreaded machine is shown in Fig. (\ref{fig:algorithm:}).
At each time step, results are accumulated for time-average integration.
The results presented on this work were carried out on a single thread.
For each classical trajectory, the procedure is repeated and the final
integration gives the spectrum intensity $I\left(E\right)$ for a
given parametric value of $E$. The same procedure is repeated for
next $E+\Delta E$, where in our calculation $\Delta E=1\mbox{cm}^{-1}$.
As previously mentioned, the trajectory is monitored by calculating
at each time step the deviation of the determinant of the monodromy
matrix from unity. The difference in the determinants was always smaller
than $10^{-6}$ during the course of the calculations. A
time step of $10$ a.u. has been always found to satisfy the strict
monodromy matrix restrictions even for the lightest atoms. 

The calculation of the full dimensional vibrational power spectrum
of the $\mathrm{CO_{2}}$ molecule is a challenging test for FP-SC-IVR
method: A successful method should reproduce spectral features such
as degenerate bending modes, strong intermodal couplings and Fermi
resonances. To evaluate the FP-SC-IVR method, we compare vibrational
spectrum of $\mathrm{CO_{2}}$ molecule from FP-SC-IVR method to numerically-exact
discrete variable representation (DVR) eigenvalue calculations on
a potential fitted to a set of first-principles points obtained at
the same level of theory. The next section describes the details of
the potential fitting and DVR calculation. Following, we continue
on the discussion of the FP-SC-IVR method.

\section{potential fitting and grid calculations}

The $CO_{2}$ molecule is a linear molecule with four vibrational
normal modes: a symmetric stretching mode ($\nu_{1}$), degenerate
bending modes ($\nu_{2}$ and $\overline{\nu}_{2}$) , and an antisymmetric
stretching mode ($\nu_{3}$). A 3\textit{d} potential energy grid
in internal coordinates is calculated using the B3LYP density functional
\citep{b3lyp} with the cc-pVDZ basis set \citep{dunningvdz}. The grid points are then fitted to a potential energy surface \citep{Requenha}
represented by a fourth-order Morse-cosine expansion\begin{align}
V\left(r_{1},r_{2},\theta\right) & =\sum_{i,j,k=0}^{4}K_{ijk}\left(1-e^{-a_{1}\left(r_{1}-r_{e}\right)}\right)^{i}\nonumber \\
 & \times\left(cos\theta-cos\theta_{e}\right)^{j}\left(1-e^{-a_{2}\left(r_{2}-r_{e}\right)}\right)^{k}\label{eq:potential_model}\end{align}
where the parameter $r_{e}=2.206119\:\mbox{a.u.}$ and $\theta_{e}=180$
specify the equilibrium coordinates of the $\mathrm{CO_{2}}$ molecule.
The Morse parameters $a_{1}=a_{2}=1.2489\:\mbox{a.u.}$ were determined
so as to minimize the standard deviation of the differences of the
fitted potential from the ab initio result using the Levenberg-Marquardt
non-linear least square algorithm \citep{fitting} . Instead,
$r_{e}$ was obtained by geometry optimization within the Q-Chem ab
initio package \citep{QChem}. 

The 35 $K_{ijk}$ coefficients were subject to the non-linear least
square fitting procedure to the DFT energies. Since these
coefficients must be the same once $r_{1}$ and $r_{2}$ are swapped,
13 linear constraints of the type $K_{ijk}=K_{kji}$ were imposed
during the fitting procedure. Additionally, to ensure that
the equilibrium geometry was fitted to the predetermined equilibrium
parametric distance, the coefficients $K_{100}$  and $K_{001}$
were constrained to be zero. Consequently, we employed a total number
of 14 fitting constraints ($K_{000}$ term is always constant).
A total of $2500$ ab initio grid points were chosen for the fitting
process. These grid points range from $1.42$ a.u. to $7.09$
a.u. for $r_{1}$ and $r_{2}$, and from $113.6$ to $180$ for
the angle variable. The calculated expansion coefficients $K_{ijk}$
are reported in Tab.(\ref{tab:K_{ijk}.}). %
\begin{table}
\begin{centering}
\begin{tabular}{cc||cc}
\hline 
{\footnotesize coeff.} & {\footnotesize attoJ} & {\footnotesize coeff.} & {\footnotesize attoJ}\tabularnewline
\hline
\hline 
{\footnotesize $K_{001}$} & {\footnotesize +0.000000} & {\footnotesize $K_{100}$} & {\footnotesize $=K_{001}$}\tabularnewline
{\footnotesize $K_{002}$} & {\footnotesize +1.442886} & {\footnotesize $K_{200}$} & {\footnotesize $=K_{002}$}\tabularnewline
{\footnotesize $K_{003}$} & {\footnotesize -0.032125} & {\footnotesize $K_{300}$} & {\footnotesize $=K_{003}$}\tabularnewline
{\footnotesize $K_{004}$} & {\footnotesize +0.003630} & {\footnotesize $K_{400}$} & {\footnotesize $=K_{004}$}\tabularnewline
{\footnotesize $K_{010}$} & {\footnotesize +0.726891} & {\footnotesize $K_{111}$} & {\footnotesize +0.392310}\tabularnewline
{\footnotesize $K_{011}$} & {\footnotesize -0.443422} & {\footnotesize $K_{110}$} & {\footnotesize $=K_{011}$}\tabularnewline
{\footnotesize $K_{012}$} & {\footnotesize -0.162970} & {\footnotesize $K_{210}$} & {\footnotesize $=K_{012}$}\tabularnewline
{\footnotesize $K_{013}$} & {\footnotesize -0.101077} & {\footnotesize $K_{310}$} & {\footnotesize $=K_{013}$}\tabularnewline
{\footnotesize $K_{020}$} & {\footnotesize +0.488451} & {\footnotesize $K_{121}$} & {\footnotesize +0.606572}\tabularnewline
{\footnotesize $K_{021}$} & {\footnotesize -0.358126} & {\footnotesize $K_{120}$} & {\footnotesize $=K_{021}$}\tabularnewline
{\footnotesize $K_{022}$} & {\footnotesize -0.210888} & {\footnotesize $K_{220}$} & {\footnotesize $=K_{022}$}\tabularnewline
{\footnotesize $K_{030}$} & {\footnotesize +0.175981} & {\footnotesize $K_{202}$} & {\footnotesize +0.097300}\tabularnewline
{\footnotesize $K_{031}$} & {\footnotesize -0.184503} & {\footnotesize $K_{130}$} & {\footnotesize $=K_{031}$}\tabularnewline
{\footnotesize $K_{112}$} & {\footnotesize +0.103205} & {\footnotesize $K_{211}$} & {\footnotesize $=K_{112}$}\tabularnewline
{\footnotesize $K_{101}$} & {\footnotesize +0.210532} & {\footnotesize $K_{040}$} & {\footnotesize +0.155374}\tabularnewline
{\footnotesize $K_{102}$} & {\footnotesize +0.067998} & {\footnotesize $K_{201}$} & {\footnotesize $=K_{102}$}\tabularnewline
{\footnotesize $K_{103}$} & {\footnotesize +0.068693} & {\footnotesize $K_{301}$} & {\footnotesize $=K_{103}$}\tabularnewline
\end{tabular}
\par\end{centering}

\caption{\label{tab:K_{ijk}.} Expansion coefficients $K_{ijk}$ for the $\mathrm{CO_{2}}$ 
B3LYP/cc-pVDZ fitted potential energy surface in attoJoule units.}

\end{table}

As far as the numerically exact eigenvalues calculations is concerned,
we used an exact DVR (Discrete Variable Representation) matrix diagonalization
procedure. The $\mathrm{CO_{2}}$ molecule was described for grid
calculations in internal coordinates, while \emph{on-the-fly} classical
trajectories and the semiclassical calculations described previously
were performed in Cartesian coordinates. No significant contamination
between the rotational (set to zero kinetic energy) and vibrational
motion was found within the simulation time. To this end, the deviation
from simplecticity of the monodromin matrix in the vibrational sub-space
were never more than $10^{-6}$ as previously mentioned.

The coordinates $r_{1}$ and $r_{2}$ are CO distances, and $\theta$
is the angle between the CO bonds. In these coordinates the kinetic
part of the Hamiltonian for $J=0$ is \begin{eqnarray}
T & = & \frac{p_{1}^{2}}{2\mu_{CO}}+\frac{p_{2}^{2}}{2\mu_{CO}}+\frac{j^{2}}{2\mu_{CO}r_{1}^{2}}+\frac{j^{2}}{2\mu_{CO}r_{2}^{2}}+\frac{p_{1}p_{2}cos\theta}{m_{C}}\nonumber \\
 &  & -\frac{p_{1}p_{\theta}}{m_{C}r_{2}}-\frac{p_{2}p_{\theta}}{m_{C}r_{2}}-\frac{cos\theta j^{2}+j^{2}cos\theta}{2m_{C}r_{1}r_{2}}\label{eq:Kinetic_internal}\end{eqnarray}
where\begin{equation}
p_{k}=-i\frac{\partial}{\partial r_{k}},\:\: k=1,2\label{eq:linear_momentum}\end{equation}
\begin{equation}
p_{\theta}=-i\frac{\partial}{\partial\theta}sin\theta\label{eq:angular_momentum}\end{equation}

and\begin{equation}
j^{2}=-\frac{1}{sin\theta}\frac{\partial}{\partial\theta}sin\theta\frac{\partial}{\partial\theta}\label{eq:j_square}\end{equation}
The carbon mass were taken to be $m_{C}=12.0$ a.m.u., while the oxygen
mass $m_{O}=15.9949$ a.m.u. and the reduced mass is as usual $1/\mu_{CO}=1/m_{C}+1/m_{O}$.

As previosuly mentioned, in order to calculate exact eigenvalues,
a sine-DVR basis for the coordinates $r_{1}$ and $r_{2}$ and a Legendre-DVR
basis for $\theta$ has been used \citep{Meyer}. For each degree
of freedom 50 DVR functions were used and eigenvalues were converged
to at least $10^{-3}cm^{-1}$. The sine-DVR ranged from $1.51$
a.u. to $3.78$ a.u. and the magnetic quantum number $m$ of the
Legendre-DVR was zero. 

Because of the restriction of total angular momentum $J=0$, we couldn't
observe all degenerate bending excitations. However, ZPE and several
vibrational energy levels were obtained and compared with that ones
coming from a single \emph{on-the-fly} semiclassical trajectory.

\section{First-Principles SC-IVR Calculations}

\begin{figure}
\begin{centering}
\includegraphics[scale=0.75]{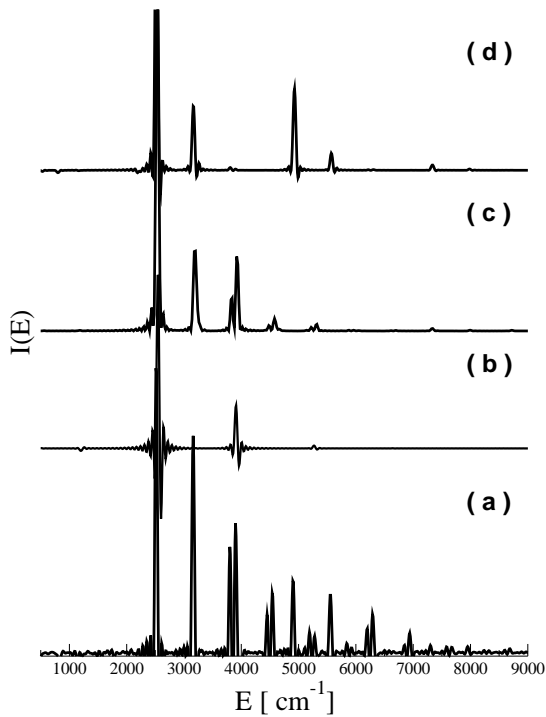}
\par\end{centering}

\caption{\label{fig:2}$CO_{2}$ Vibrational Power Spectrum: Initial kinetic
energy on: (a) all modes; (b) symmetric mode; (c) one bending and
symmetric modes; (d) bending and asymmetric modes.}

\end{figure}

The full power spectrum obtained using Eq. (\ref{eq:TA_spectrdens})
after 3000 BOMD steps of 10 a.u. each is shown on the bottom of Fig.
\ref{fig:2}.%
 For longer simulations, the monodromy matrix symplectic properties
as well as the resolution of the spectrum started to deteriorate.
The calculated vibrational zero-point energy (ZPE) value was $2518\:\mbox{cm}^{-1}$
versus the exact value of $2514.27\:\mbox{cm}^{-1}$ and both are
in good agreement with the experimental value of $2508\:\mbox{cm}^{-1}$.
In contrast, harmonic normal-mode analysis (whose frequencies are
$656.62,\:1363.46,\:2423.47$ wavenumbers) predicts a frequency of
$2550.08\:\mbox{cm}^{-1}$. Thus, the TA-SC-IVR method successfully
reproduces the ZPE anharmonic effects with the use of a single classical
trajectory. Some representative frequencies of the power spectrum
are presented in Table \ref{tab:Eigenvalues}. The ZPE was shifted
to zero for comparison with reported classical ELMD simulations on
the same system that cannot reproduce the ZPE or higher vibrational
states \citep{Gygi,Saad} but only single modes frequencies. For these
studies of Refs. \citep{Gygi,Saad}, the vibrational data were obtained
from the Fourier transform of correlation functions of classical trajectories
in plane-wave DFT calculations. The ELMD approach predicts the following
fundamental frequencies $648,\:1368,\:1428$ and $2353$ for Ref.
\citep{Gygi} and $663,\:1379,\:1456$ and $2355$ for Ref. \citep{Saad}.
These classical results are similar but limited to a normal mode analysis.

Table \ref{tab:Eigenvalues} compares our TA-SC-IVR results with the
exact ones and to those obtained by Filho \citep{Filho} with the
same density functional and a basis set of comparable quality (6-31+G{*})
\citep{pople631g}, using a perturbative approximation of the eigenvalue
expansion. One can see how a different basis set results a significant
deviation of vibrational levels spacing, once the comparison is performed
in units of wavenumbers. %
\begin{table}
\begin{minipage}[t][1\totalheight]{1\columnwidth}%
\begin{center}
{\footnotesize }\begin{tabular}{cccccc}
\hline 
{\footnotesize Exp.}%
\footnote{Experimental frequencies in $\mathrm{cm^{-1}}$ from Ref. \citep{BrownFarmer}%
} & {\footnotesize mode}%
\footnote{First number is the symmetric stretch quantum, second are the degenerate
bendings, and third one is the asymmetric stretch. The exponent of
the second number is the $l_{i}$ degeneracy index.%
} & {\footnotesize Harmonic}%
\footnote{Vibrational levels according to a normal modes harmonic model%
} & {\footnotesize FP-SCIVR-SA}%
\footnote{Using the Separable approximation of Eq.(\ref{eq:sep_approx})%
} & DVR & {\footnotesize Ref.} \citep{Filho}\tabularnewline
\hline
\hline 
{\footnotesize 667.4} & {\footnotesize $0,1^{1},0$} & {\footnotesize 656.62} & {\footnotesize 644} &  & {\footnotesize 657.2}\tabularnewline
{\footnotesize 1285.4}$^{\wedge}$ & {\footnotesize $0,2^{0},0$} & {\footnotesize 1313.24} & {\footnotesize 1288} & {\footnotesize 1252.91} & {\footnotesize 1283.4}\tabularnewline
{\footnotesize 1388.2}$^{\wedge}$ & {\footnotesize $1,0^{0},0$} & {\footnotesize 1363.46} & {\footnotesize 1381} & {\footnotesize 1372.29} & {\footnotesize 1408.8}\tabularnewline
{\footnotesize 1932.5}$^{\dagger}$ & {\footnotesize $0,3^{1},0$} & {\footnotesize 1969.86} & {\footnotesize 1932} &  & {\footnotesize 1930.2}\tabularnewline
{\footnotesize 2003.2} & {\footnotesize $0,3^{3},0$} & {\footnotesize 1969.86} & {\footnotesize 2024} &  & {\footnotesize 2004.9}\tabularnewline
{\footnotesize 2076.9}$^{\dagger}$ & {\footnotesize $1,1^{1},0$} & {\footnotesize 2020.08} & {\footnotesize 2106} &  & {\footnotesize 2098.5}\tabularnewline
{\footnotesize 2349.1} & {\footnotesize $0,0^{0},1$} & {\footnotesize 2423.47} & {\footnotesize 2388} & {\footnotesize 2359.51} & {\footnotesize 2411.5}\tabularnewline
{\footnotesize 2548.4}$^{\ddagger}$ & {\footnotesize $0,4^{0},0$} & {\footnotesize 2626.48} & {\footnotesize 2515} & {\footnotesize 2482.95} & {\footnotesize 2553.3}\tabularnewline
{\footnotesize 2585.0}$^{\star}$ & {\footnotesize $0,4^{2},0$} & {\footnotesize 2626.48} & {\footnotesize 2578} &  & {\footnotesize 2591.2}\tabularnewline
{\footnotesize 2671.7}$^{\ddagger}$ & {\footnotesize $0,4^{4},0$} & {\footnotesize 2626.48} & {\footnotesize 2669} & {\footnotesize 2640.15} & {\footnotesize 2716.5}\tabularnewline
{\footnotesize 2760.7}$^{\star}$ & {\footnotesize $1,2^{2},0$} & {\footnotesize 2676.70} & {\footnotesize 2759} &  & {\footnotesize 2796.3}\tabularnewline
{\footnotesize 2797.2}$^{\ddagger}$ & {\footnotesize $2,0^{0},0$} & {\footnotesize 2726.92} & {\footnotesize 2793} & {\footnotesize 2757.14} & {\footnotesize 2845.2}\tabularnewline
{\footnotesize 4673.3} & {\footnotesize $0,0^{0},2$} & {\footnotesize 4846.94} & {\footnotesize 4690}$^{+}$ & {\footnotesize 4693.24} & {\footnotesize 4797.8}\tabularnewline
{\footnotesize 6972.6} & {\footnotesize $0,0^{0},3$} & {\footnotesize 7270.41} & {\footnotesize 6803}$^{+}$ & {\footnotesize 6821.35} & {\footnotesize 7152.9}\tabularnewline
\hline
\end{tabular}
\par\end{center}%
\end{minipage}%

\caption{\label{tab:Eigenvalues}Some of the calculated vibrational energy
eigenvalues. All data are in wavenumbers. Fermi Resonances group of
frequencies are indicated by the same superscript symbols. Uncertain
peaks are marked with $\left(+\right)$. The first column represents
the experimental vibrational frequencies associated with the modes
listed on the second column. The third column shows the harmonic DFT
results. In the fourth and fifth columns, we show our FP-SCIVR and
exact numerical DVR calculations in the B3LYP/cc-PVDZ model chemistry
used for the FP-SCIVR calculations. The fifth column shows perturbative
DFT calculations carried out using a similar functional and basis
set.}

\end{table}

A major difficult on the $\mathrm{CO_{2}}$ power spectrum simulations
is the calculation of the Fermi resonance splittings. These are the
result of anharmonic couplings, and they represent a stringent test
for a semi-classical method that relies on a single short trajectory.
The Fermi resonances occur when an accidental degeneracy between two
excited vibrational levels of the same symmetry exists and it results
in a repulsion between the corresponding energy levels. The sources
of these resonances are purely anharmonic and are only present in
polyatomic potentials. For the $\mathrm{CO_{2}}$ molecule, the unperturbed
frequencies for the symmetric stretching are roughly equal to the
first bending overtone ($\nu_{1}\cong{2\nu}_{2}$). For these modes,
the wavefunctions are transformed as the irreducible representation
of $D_{\infty h}$, \emph{i.e}. $\nu_{1}$$\left(10^{0}0\right)$
as $\Sigma_{g}^{+}$, at the experimental frequency of $1388\:\mbox{cm}^{-1}$,
and $\nu_{2}^{2}$$\left(02^{0}0\right)$ as $\Sigma_{g}^{+}+\Delta_{g}$,
at an experimental frequency of $1285\:\mbox{cm}^{-1}$. Another Fermi
doublet results from the addition of a quantum of bending mode to
the previous Fermi doublet to yield the following states: $\nu_{1}\nu_{2}$$\left(11^{1}0\right)$
, at an experimental frequency of $2077\:\mbox{cm}^{-1}$ and the
$\nu_{2}^{3}$$\left(03^{1}0\right)$ state, at an experimental frequency
of $1932\:\mbox{cm}^{-1}$. Higher-energy Fermi resonances are indicated
in Table \ref{tab:Eigenvalues} by using the same superscript symbols.
The first Fermi terms are located at $1313$ and $1363$ in a harmonic
approximation and corrected to $1288$ and $1381$ wavenumbers for
FP-TA-SC-IVR. Thus, the original levels have been repelled by Fermi
couplings. One mode is located at a higher frequency than the harmonic
prediction, while the other is at a lower frequency. The latter effect
could be explained also by simple anharmonicity, but the former is
evidence of the ability of the single trajectory FP-TA-SC-IVR method
even when the separable approximation is used to capture Fermi resonance
effects partially. The same reasoning can explain the second Fermi
doublet located at $1932$ and $2106$ for FP-TA-SC-IVR, while the
harmonic estimate at $1970$ and $2020$ wavenumbers. 

With the FP-TA-SC-IVR method, one can also identify the couplings
between vibrational modes and the appearance of Fermi resonance splittings
by carrying out simulations with different initial conditions. This
can be achieved by selectively setting the initial velocity of some
vibrational modes to zero. The anharmonic coupling between levels
leads to a consistent reproduction of the ZPE peak in the spectrum
for all simulations. However the excited vibrational peaks related
to the modes with zero initial kinetic energy show a very small signal
in the power spectrum. Vibrational energy redistribution processes
can be studied as well, by carrying out simulations at different
timescales. In Fig. \ref{fig:2}, we show the resulting power spectra
for different initial conditions. If the initial state contains only
purely symmetric motion, the lowest Fermi resonance peaks in Fig.
\ref{fig:2}(b) are absent as well as for a bending (without symmetric
stretching) motion in Fig. \ref{fig:2}(d). These results and the
intensity of their peaks respect to that ones located at the same
frequencies in Fig. \ref{fig:2}(a) suggest that the Fermi resonance
is indeed originated from the coupling between bending and the symmetric
modes. One can reach the same conclusions by inspecting the lower
Fermi doublet peaks intensity: by adding a bending mode (from Fig.
\ref{fig:2}(b) to Fig. \ref{fig:2}(c)) and a second one (from Fig.
\ref{fig:2}(c) to Fig. \ref{fig:2}(a)) the intensity of both peaks
is gradually raised. This is called {}``intensity borrowing'' and
it arises from the strong mixing of the zero order states. These observations
reinstate that {}``repulsion and mixing are the hallmarks of Fermi
resonances'' \cite{Heller_fermiCO2}. Also, for a distinct
set of initial conditions, an additional peak at $5500$ cm$^{-1}$
related to the asymmetric stretch was observed. Using the proposed
approach, one can carefully detect the characteristics of each peak
even for complicated power spectra.%
\begin{figure}
\begin{centering}
\includegraphics[scale=0.75]{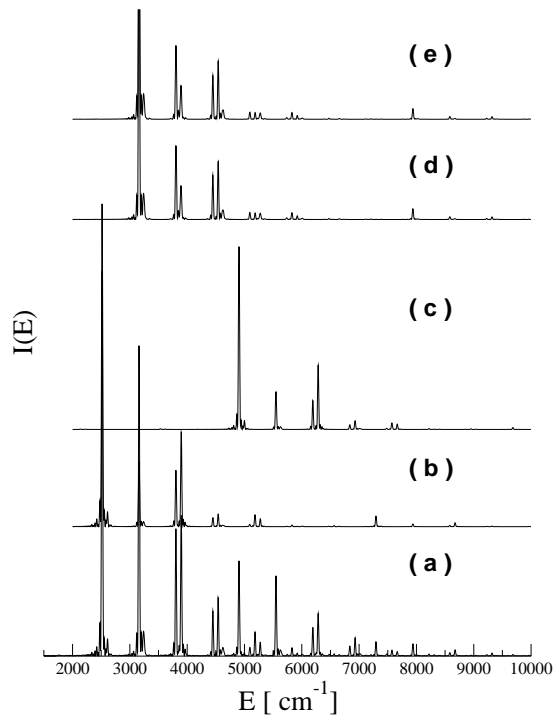}
\par\end{centering}

\caption{\label{fig:3}$CO_{2}$ Vibrational Power Spectrum (Separable approximation):
Different basis set symmetries for $\nu_{1}$(symmetric stretching
mode), $\nu_{2}$ and $\overline{\nu}_{2}$ (bending modes) and $\nu_{3}$(asymmetric
mode) and the corresponding $D_{2h}$ irreducible representation;
(a) all $\epsilon$s are zero; (b) $\left(B_{1u}\right)$: $\epsilon\left(v_{1}\right)=0,\epsilon\left(v_{2}\right)=1,\epsilon\left(\overline{\nu}_{2}\right)=0,\epsilon\left(v_{3}\right)=-1$;
(c) ($A_{g}$): $\epsilon\left(v_{1}\right)=1,\epsilon\left(v_{2}\right)=0,\epsilon\left(\overline{\nu}_{2}\right)=0,\epsilon\left(v_{3}\right)=1$;
(d) ($B_{2u})$:$\epsilon\left(v_{1}\right)=0,\epsilon\left(v_{2}\right)=-1,\epsilon\left(\overline{\nu}_{2}\right)=0,\epsilon\left(v_{3}\right)=1$,
(e) $\left(B_{3u}\right)$ $\epsilon\left(v_{1}\right)=0,\epsilon\left(v_{2}\right)=0,\epsilon\left(\overline{\nu}_{2}\right)=-1,\epsilon\left(v_{3}\right)=1$.
$B_{2u}$ and $B_{3u}$ representations are degenerated in the $D_{\infty h}$
subspace as shown.}

\end{figure}

An attractive method for obtaining the symmetry properties of the
eigenstates involves arranging the initial basis vectors \citep{SunMiller_HCldimer,Alex_Mik}.
The basis for this method is the direct product of coherent states
$\vert\chi\rangle=\prod_{k=1}^{4}\vert p_{i}^{\left(k\right)},q_{i}^{\left(k\right)}\rangle^{\epsilon_{k}}$
. These states can be chosen to have an initial symmetry by employing
linear combinations of the form $\vert p_{i}^{\left(k\right)},q_{i}^{\left(k\right)}\rangle^{\epsilon_{k}}=\left(\vert p_{i}^{\left(k\right)},q_{i}^{\left(k\right)}\rangle+\epsilon_{k}\vert{-p},-q_{i}^{\left(k\right)}\rangle\right)/\sqrt{2}$.
The $k$-th mode can be made symmetric $(\epsilon_{k}=1$), antisymmetric
($\epsilon_{k}=-1$) or have no symmetry restrictions ($\epsilon_{k}=0$).
In order to assign the proper symmetry to each peak on Fig. \ref{fig:3}
, the reduced $D_{2h}$ symmetry group was adopted. All irreducible
representations were reproduced and peaks were grouped by symmetry
as reported in Fig. \ref{fig:3}. Note that (d) and (e) plots are
identical since they only differ trivially by swapping coefficients
between the degenerate bending modes in the original $D_{\infty h}$
symmetry group.%
\begin{figure}
\begin{centering}
\includegraphics[scale=0.75]{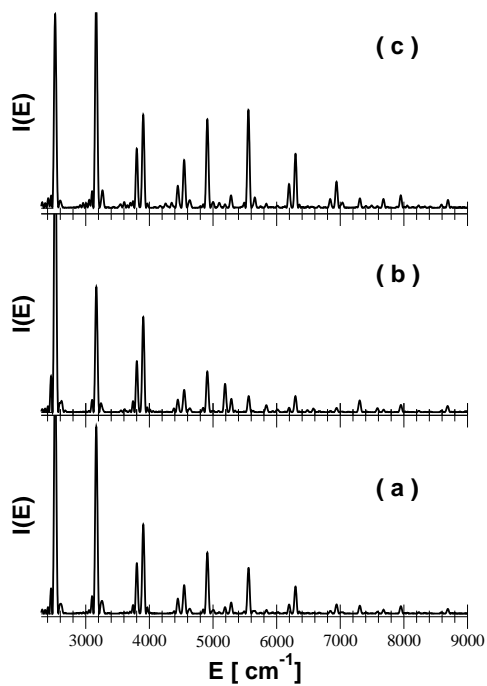}
\par\end{centering}

\caption{\label{fig:Gaussian-width}Gaussian width variations and related power
spectra: a) $\gamma_{i}=\omega_{i}$; b)$\gamma_{i}=2\omega_{i}$;
c)$\gamma_{i}=\omega_{i}/2$, where $\omega_{i}$ are the $i-esime$
normal mode frequency. The FP-SCIVR power spectra are fairly insensitive
to variations in the value of the coherent state width.}

\end{figure}

Finally we have investigated the stability of the propagator versus
variations of the coherent states gaussian width parameters $\gamma_{i}$.
Previous calculations \citep{HermanLukcoherstates} have shown that
there is no significant depedency on energy and norm conservation
for the semiclassical propagator if suitable values of $\gamma_{i}$
are chosen. For power spectra calculation we have chosen to look at
vibrational levels variations under different values of coeherent
states width. Since a single trajectory was used in the FP-TA-SC-IVR
approach, no Monte Carlo integration is performed in phase space coordinates
and the changes of $\gamma_{i}$ are confined to the coherent states
overlap and to the prefactor in Eq. (\ref{eq:prefactor}). As reported
in Fig. \ref{fig:Gaussian-width} and checked on a finer scale, no
significant variation was observed beyond $1\:\mbox{cm}^{-1}$. These
findings are in agreements with previous calculations on the same
propagator \citep{HermanLukcoherstates}. Interestingly,
a different distribution in peaks intensity were found in each panel.
Since the peaks magnitude is proportional to the overlap between the
reference state and the actual eigenfunction, the anharmonic choice
($\gamma_{i}=\omega_{i}/2$) is a more suitable solution as clearly
showed on panel (c) of Fig. \ref{fig:Gaussian-width}.

\section{Conclusions }

\begin{figure}
\begin{centering}
\includegraphics[scale=0.75]{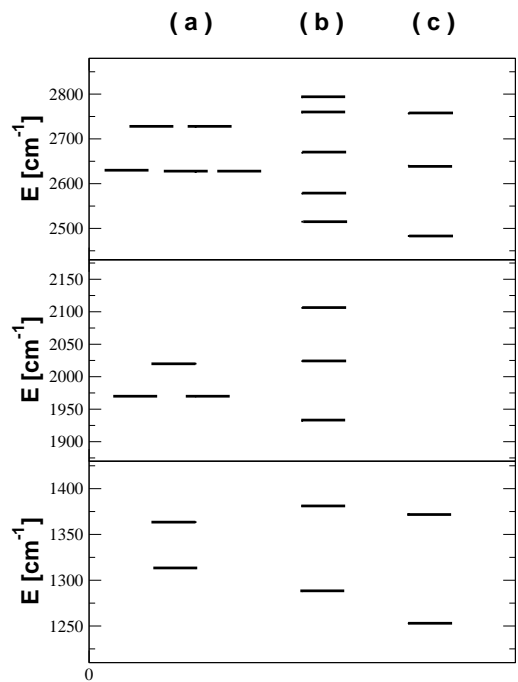}
\par\end{centering}

\caption{\label{fig:5}Fermi Resonance states vibrational energy level: (a)
in harmonic approximation; (b) single FP-SC-IVR trajectory calculation;
(c) exact grid calculation on splined potential.}

\end{figure}

In conclusion, we have shown that SC-IVR can be implemented easily
and efficiently using first principles molecular dynamics. With the
modest computational cost of a single classical trajectory, the vibrational
density of states of the $\mathrm{CO_{2}}$ molecule was calculated.
On Fig. \ref{fig:5} we report a graphical comparison between the
harmonic and the FP-TA-SC-IVR approximations, versus the exact vibrational
value for the Fermi resonance multiplets. One can notice how the single
trajectory FP-TA-SC-IVR goes far beyond the harmonic approximation
by removing the harmonic degenerancy and including part of anharmonicity.
Fermi splittings are well mimiced not only for the first doublet,
but also for the higher ones. The numerically exact DVR vibrational
energy levels constrained by $J=0$ are represented on the last column.
The FP-TA-SC-IVR values are similar to the DVR results, when comparison
is possible. However, a closer look at Table (\ref{tab:Eigenvalues})
shows how these single trajectory FP-TA-SC-IVR calculations can include
only part of the anharmonicity and that their precision gets worse
for higher vibrational levels.%
In particular, the spacing of the higher-energy states is
harmonic-like and thisis the mayor  limitation of using a single classical
trajectory. 

These and previous calculations on model potentials \citep{Alex_Mik}
has shown how the single trajectory TA-SC-IVR gives reasonable results
and performs better for higher frequencies modes. The computational
cost of the method is essentially the same as classical propagation,
and therefore, if broadly implemented in electronic structure codes,
it can provide a description of quantum effects at a comparable computational
cost to that of classical approaches. Possible applications of this
method or related ones are the study of excited electronic states
and Franck-Condon transitions, such as vibrational absorption spectra
\cite{Jorge}. Although this single trajectory approach may be a practical
tool for the simulation of more complex systems, the use of more trajectories
is probably required to remove any harmonic``ghost states''.
We are currently exploring the use of a small number of a set of systematically
determined trajectories for further improvement of the results. If
the number of required trajectories grows as a low polynomial of the
system size, semi-classical methods could be competitive with currently-employed
numerical approximations to obtain anharmonic vibrational effects.
Finally, we expect that the representation of the potential energy
in terms of normal coordinates will become less suitable when large
amplitude motions or non adiabatic effects come into play.

\section*{Acknowledgement}

One of the authors (M.C.) feels deeply in debt with Prof. W. H. Miller
for the many lessons learned from him. The authors thank Dr. A. Kaledin,
Prof. E. J. Heller for useful discussions and revision of the manuscript.
A. A.-G., S. S. and S. A. thank the Faculty of Arts and Sciences of
Harvard University for financial support and S. S. thanks the Samsung
Scholarship for financial support. M. C. and G. F. T. thanks the University
of Milan for fundings and CILEA (Consorzio Interuniversitario Lombardo
per L'Elaborazione Automatica) for computational time allocation.
A. A.-G., S. S. and S.A thank FAS Research Computing for cluster computing
support.

\end{document}